# UPS and UV spectroscopies combined to position energy levels of $TiO_2$ anatase and rutile nanopowders.

Clément Maheu*, Luis Cardenas*, Eric Puzenat, Pavel Afanasiev, Christophe Geantet

Université de Lyon, Institut de Recherches sur laCatalyse et l'Environnement de Lyon -IRCELYON -UMR 5256, CNRS-UCB Lyon 1,2 avenue Albert Einstein 69626 Villeurbannecedex, France

* corresponding author: clement.maheu@ircelyon.univ-lyon1.fr, luis.cardenas@ircelyon.univ-lyon1.fr

## Abstract

An accurate experimental determination of electronic structures in semi-conductor nanopowders is a challenging task. We propose here to combine UPS and UV-Vis spectroscopies in order to get the full description of the electronic band alignment of powder samples, $TiO_2$ rutile and anatase. For UPS measurements, two preparation methods, namely the dropping method and electrophoretic deposition, were used to prepare layers of titania powders on a conducting substrate, ITO or Ag. Both methods lead to comparable results, with a quantitative description of the energy levels from the valence band. Combining these results with the UV-Vis spectra of the same powders enables the determination of the absolute position of the valence band maximum and the conduction band minimum. Combined UPS-UV-Vis spectroscopy provides a better insight into the properties of a powdered material which can differ from single crystal model systems. It can also be used to predict the electronic transfer in mixed phase systems during photocatalytic processes.

# Introduction

During the last decades, extensive efforts were made to determine precisely the energy levels and band structure of titania ($TiO_2$).[1] Titanium oxide is used in many areas as memresistors, optoelectronics, photovoltaics or photocatalysis.[2–4] Determination of electronic structures and carriers transport properties is crucial for these applications. Regarding photocatalysis, titania powders is the most widely studied class of materials to the present day.[5] In the 70s, Teichner discovered and conceptualized heterogeneous photocatalysis with $TiO_2$.[6,7] Since, a tremendous amount of studies have been devoted to this oxide and its photocatalytic properties.[7,8]

Among the many polymorphs of $TiO_2$, rutile and anatase are the most studied.[2] For the photocatalytic applications of these crystalline structures, determination of the position of the energy levels is essential. Position of the valence band (VB), of the conduction band (CB) and the size of the optical band gap ($E_g$) are dependent of the crystalline phase. These parameters influence the amount of photons absorbed upon illumination and the respective oxidation and reduction potentials of holes and electrons, generated by light. Presence of defects or dopants as well as the particle size may also play a key role and allow tuning the electronic properties of titania materials.[10]

The relative positions of the energy levels determine the direction of the electronic transfer of a photocatalytic material based on a heterojunction between $TiO_2$ and a metal[11] or a semiconductor.[12] A particular interest is also devoted to the materials composed of different titania phases. The most known one is Degussa P25 titania which is a mixture of rutile and anatase.[13] The exact nature of its high photocatalytic activity still remains not completely understood, but the energy diagrams are obviously crucial to explain its properties.[14] The bands alignment between the two phases in Degussa P25 was claimed to enhance the charge careers separation.[15] However, there is a controversy in the literature concerning this band alignment. The band gaps of rutile and anatase $TiO_2$ polymorphs are of 3.0 and 3.2 eV, respectively. However, their relative position is still under debate. Three possibilities are generally discussed (Figure 1). The energy levels of both phases can be included as the $E_{g,rutile} < E_{g,anatase}$ (type I).[15] It can also be staggered alignment with the CB and the VB of anatase lying above those of rutile (type II-anatase) or with the CB and the VB of rutile lying above those of anatase (type II-rutile).[16] As the bands are close, Mi *et al.* proposed two additional particular cases with the VB or the CB of the two phases being aligned at the same value.[17] However, there are no physical methods accurate enough to prove the exact match of the energy levels.

The band positioning of type II-rutile or type I should induce an electronic transfer from anatase to rutile.[18–21] For type II-anatase, the charge transfer is in the opposite direction.[16,22–24]

The first paper to advocate a type II-rutile alignment came from Mott–Schottky analysis by Kavan et al.[18] This hypothesis was supported by experimental data[19,21] and calculations.[20] However, recently it was shown that the Mott–Schottky analysis is strongly influenced by the surface charges.[16,25] Adsorption of ions at the surface of $TiO_2$ leads to the appearance of a dipole layer which interferes with band alignment analysis provided by this technique.

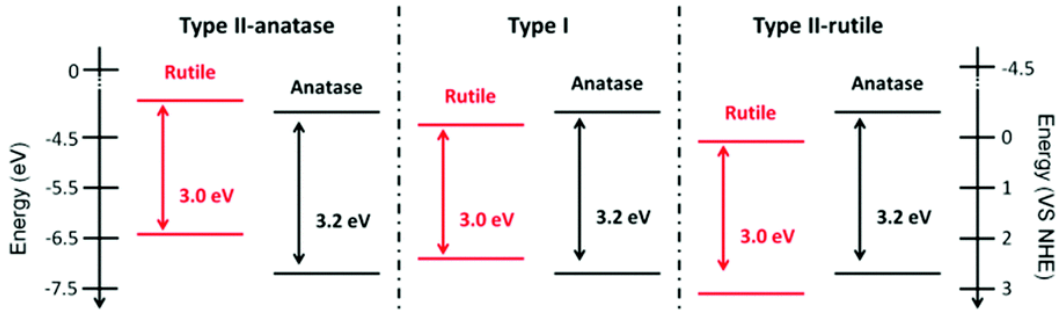

**Figure 1:** Possible cases of relative positions of rutile and anatase energy levels. Type I, included; type II, staggered with energy levels of anatase above or reverse case.

The hypothesis of the type II-anatase band alignment was mainly supported by XPS measurements.[23,24] The offset between the $E_{VB}$ of two phases was measured and then, the $E_{CB}$ offset is deduced from the known $E_g$ difference. Scanlon et al. reported a difference of 0.4 eV[23] and Pfeifer et al. of 0.6–0.7 eV.[24] Gao and co-workers claimed a similar band alignment using Kelvin probe force microscopy. This work has the particularity to evidence the band alignment for a real heterojunction, previously it was shown only on separate $TiO_2$ phases.[26]

Recent theoretical experiments support the type II-anatase band alignment but with a higher band offset.[27] Ko et al. using the relativistic DFT calculation suggested that the type of alignment depends on the size of rutile and anatase particles.[28] It means that sizes should be taken into account as well as the presence of surface states, impurities or structural defects. Indeed, between two powders, even within the same phase, these parameters can induce variation of the electronic structure. Therefore, clear and accurate experimental protocols are needed to determine the absolute electronic structure of any $TiO_2$-based powder materials (extendable to any powder materials). To this day, no such protocol is reported in the literature. Two spectroscopic techniques were used, UV-Vis Diffuse Reflectance Spectroscopy (DRS) to study transitions from the VB to the CB and Ultraviolet Photoelectron Spectroscopy (UPS).

The latter is a surface sensitive tool, widely used to probe the electronic valence band.[29,30] UPS provides information on the distribution of electronic density in the VB as well as on the work function. UPS has been extensively applied to undoped and doped model titania surfaces.[2,31,32] The UPS results obtained on the $TiO_2$(001) monocrystalline surface highlight the formation of new structures with narrower band gaps after reductive or oxidative treatments[33,34] or after nitrogen implantation.[35] Formation of new electronic states located above the O 2p band after N-doping has been evidenced with this method. As another example, UPS on $VO_x/TiO_2$ (001) showed the presence of intermediate states near the Fermi level.[36]

However, the most efficient $TiO_2$ photocatalysts are not single crystals but nanopowders. Morphological and chemical differences may induce differences between the model materials, prepared under a controlled atmosphere and with a true electronic structure of powdered samples.

In this paper we introduce a new methodology to study photocatalytic powders, in our case titania, with UPS and provide a combined UPS-UV-Vis spectroscopic study of rutile and anatase polymorphs in order to establish the absolute energy band positions.

# Experimental section

## Materials

The two $TiO_2$ polymorphs studied are commercial powders. Anatase was purchased from Cristal (CristalACTiV™ PC500) and rutile was purchased from US Research Nanomaterials (99.9% – 30 nm). The XRD patterns were recorded on a Bruker D8 Advance A25 diffractometer equipped with a Ni filter (Cu Kα radiation at 0.154184 nm).

The crystallite size was estimated from X-ray diffraction (XRD) line broadening using the Scherrer equation and found to be around 7 nm for anatase and 25 nm for rutile as detailed in the Supp. Info., Figure S1.

## Ultraviolet photoelectron spectroscopy

UPS measures the kinetic energy of photoelectrons emitted by solid surfaces upon irradiation with ultraviolet light, in order to determine the electronic structure of the valence band.[37] UPS analyses were conducted in an integrated ultrahigh vacuum system, connected to an A Axis Ultra DLD spectrometer (Kratos Analytical). UPS spectra were obtained with a He lamp providing two resonance lines, He I (hν = 21.2 eV) and He II (hν = 40.8 eV). To separate the secondary edges of the sample and analyzer, a negative potential of about 9.1 V was applied to the sample. Prior to measurements, an Ag foil was cleaned by argon ion sputtering. The bias was optimized to shift the Ag spectrum in the linear region of the analyzer (0–10 eV kinetic energy). A work function of 4.2 eV was obtained, in good agreement with the literature value of around 4.26 eV (see the Supp. Info.).[38] The spectrometer has a pass energy of 10 eV and a spatial aperture of 110 µm. Under these conditions, the intrinsic resolution of the spectra was measured at the Fermi edge of the Ag foil at room temperature (see Figure S2), as reported by Schalf et al.[39] It is estimated to be 0.15 eV. For comparison, the intrinsic resolution of a lab-based XPS setup using a monochromatized source is more than 0.5 eV.

At high binding energy (BE), the cut-off energy ($E_{cut-off}$) is determined using the first derivative of the UPS spectrum, as detailed in the Supp. Info. At lower BE a strong contribution of the secondary electrons required a careful background subtraction. In the literature, different types of background are used for UPS spectra interpretation. Fleming et al. advised an "a + cxd" form,[40] Laidani and co-workers recommended a Shirley type background[41–43] and Payne et al. used a Tougaard type background.[44] These types of background are more suitable for higher photon energies. Li et al. proposed a background dedicated to UPS analysis.[45] It takes into account the energy dependence of the cross-section for the creation of secondary electrons by primary electrons. It also considers the secondary-electron cascade process, i.e. those secondary electrons can in turn generate other secondary electrons. The latter induces an exponential shape which is not represented by a standard core-level background. Comparison of the Li background with a core-level type background, Figure S4, advocates the use of the Li background. After background subtraction, the signal was decomposed using Lorentzian Asymmetric Finite (LF) function.

To determine the valence band maximum ($E_{VB}$), we proposed to use the second derivative of the spectrum acquired at low binding energy. $E_{VB}$ is the energy at which the first electron is photogenerated. Therefore, the $E_{VB}$ and the inflexion point should match. It provides similar results to the method proposed by Sayan et al.[46]

## UPS sample preparation

Two different types of substrates and two sample preparation methods were used. The first substrate is a commercial conducting ITO layer on glass with $R_S$ = 30–60 Ω sq.$^{-1}$, purchased from Sigma. The second is an Ag foil purchased from Alfa Aesar, previously cleaned by Ar ion sputtering. The UPS spectra of both substrates were acquired as references (Figures S2 and S3). The $E_{cut-off}$ measured to be 17.0 eV for ITO agrees with that reported by Beerbom et al.[47] It confirms the accurate calibration of the analytic system. ITO is a suitable system for UPS measurements since there is no major electronic contribution in the range of 10–0 eV which is our area of interest. In contrast, the d bands of Ag are located in this energy range, but the Ag foil permits an absolute measurement of $E_{VB}$.

The samples for UPS measurements must be thin in order to minimize the charging effects. For the first preparation method, the powders were ultrasonically dispersed in heptane or pentane (4 g L$^{-1}$). Then, several drops were deposited on the conducting ITO substrate. This preparation method was proposed for zeolites[48] and further applied to other oxides such as $V_2O_5$ or bismuth molybdate.[49] In the second preparation method, a homogeneous thin film of the $TiO_2$ powder was prepared by electrophoretic deposition (EPD). This technique was first used to prepare samples of dye-sensitized solar cells.[50] The particles of a stable suspension were deposited by applying a DC electric field.[51] The EPD duration, electrode distances, and composition of the solution can influence the thickness and the morphology of the film.[52] Because of the key role of the charging solution, several solvents were tested.[53–55] One of the most efficient systems is a mixture of iodine, acetone and water.[56] Therefore, the EPD technique enables the preparation of a homogeneous, pure and well-controlled $TiO_2$ layer on a conducting substrate.

To prepare the samples by EPD, 50 mg of iodine was dissolved in 100 mL acetone upon stirring. Then, a suspension of titania powder (2–3 g L$^{-1}$ concentration) was added, the suspension was sonicated for 10 min and left for approximately 1 h. It was poured in a glass vessel containing two parallel ITO/glass substrate bars, separated by ca. 1 cm. Then, a constant voltage of 14.9 V was applied for 1 min with a stabilized power supply (type elc AL 936N).

## UV-Vis absorption spectroscopy

UV-Vis diffuse reflectance was measured at room temperature with a Perkin Elmer Lambda 45 spectrometer equipped with an integrating sphere (RSA-PE-20). Pure $BaSO_4$ was used as a reference and to dilute the $TiO_2$ powder. It has a total reflection of the light in the wavelength range of interest of 200–700 nm. If the sample is thick enough (here 4–5 mm), no light is transmitted and the Kubelka–Munk (KM) function $F_{KM}$ can be applied. The KM equation is valid for weakly absorbing materials, whereas for the values of reflectance $R_\infty$ < 0.6, deviations from the theory occur. Therefore, highly diluted samples are required for eliminating the specular reflection and minimizing the grain size effects.[57] To prepare highly diluted samples, the standard procedure was to mix 30 mg of the $TiO_2$ powder with 1200 mg of $BaSO_4$ (a mass ratio of 1:40). Then 30 mg of the latter mixture was again mixed with 1200 mg of $BaSO_4$ to get a 1:1600 ratio (625 ppm). This iterative procedure was repeated to get samples with a mass weight of 0.39 ppm. Such a procedure allows obtaining high quality UV-Vis spectra.

Additionally, dilution of the samples permits approximation of the scattering coefficient by that of BaSO$_4$. The absorption coefficient is then easily obtained from reflectance measurement (see the Supp. Info.). The latter is used to obtain the band gap of both anatase and rutile using Tauc's relationship[58] as recommended by López et al.[59] A further decomposition of a UV spectrum ($F_{KM}(hv)$) can be observed after normalization according to the equation:

$$t(hv) = \frac{\int_{hv-dhv}^{hv+dhv} F_{KM}(hv)dhv}{\int_0^\infty F_{KM}(hv)dhv}$$

This new t(hv) function can be decomposed with several Gaussian contributions. Details of the fitting parameters are reported in the Supp. Info.

# Results and discussion

## Valence band maximum ($E_{VB}$): UPS

While UPS allows getting a higher resolution and better sensitivity than XPS, the sample preparation in the case of non-conductive powders is more demanding as concerns the layer homogeneity and its electric contact with the conducting support. Two different preparation methods of samples for UPS measurements are compared in the present study. In the first approach, a suspension of $TiO_2$-anatase in alkanes was dropped on the ITO substrate. Pentane and heptane suspensions were studied. As UPS is sensitive to impurities, the first has the advantage of being easily removed. In the second approach $TiO_2$-anatase was deposited by electrophoresis. All the UPS spectra are reported in the Supp. Info. They have a similar shape at low BE, as the occupied bonding and non-bonding states due to $TiO_2$ are observed.[2,57] At higher BE, the secondary edge is observed and corresponds to the $E_{cut-off}$.

For the spectra shown in the Supp. Info. (Figures S5 and S6), anatase $E_{VB}$ values are similar. $E_{VB}$ varies from 2.9 to 3.1 eV; the $E_{cut-off}$ differs from one preparation type to another. The presence of an interfacial dipole between the ITO substrate and the $TiO_2$ powder may induce a global shift of the spectrum. This dipole is influenced by the preparation method. A method to confirm the absolute position of $E_{cut-off}$ and $E_{VB}$ will be discussed below.

UPS spectra obtained for anatase and rutile are shown in Figure 2. The $E_{cut-off}$ is measured for both polymorphs to be 17.2 eV but a shift is observed at low energy. The measured $E_{VB}$ values are 2.9 eV for anatase and 2.5 eV for rutile. This difference of 0.4 ± 0.2 eV agrees with the previous result reported by Scanlon et al. and Pfeifer et al.[23,24] However, it is the first time that this difference is measured on powdered titania with the UPS method. Note that the same band offset between the two polymorphs was obtained with the method determining $E_{VB}$ as the intersection point between the zero-intensity constant backgrounds and a linear extrapolation of the valence band decay (Table S1).[61,62]

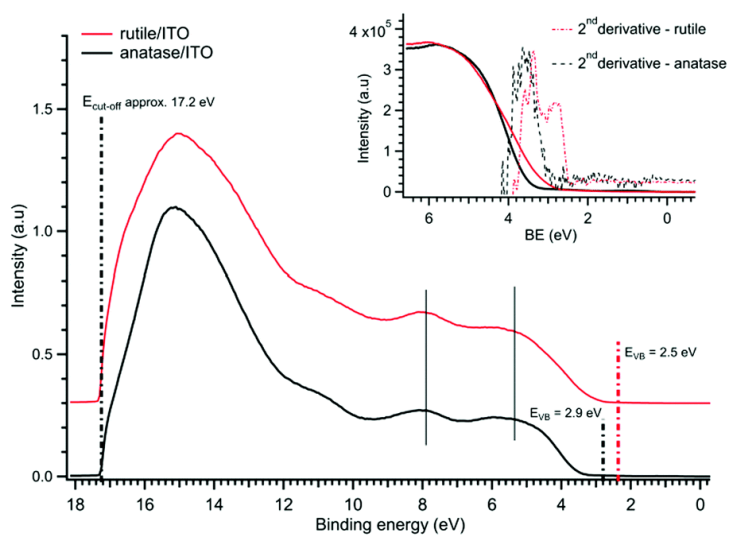

**Figure 2:** He I UP spectra of anatase/ITO (black) and rutile/ITO (red); both prepared by the EPD method. Inset – Zoom on the low BE area for anatase (black) and rutile (red). Black and red dotted lines correspond to the second derivative of the corresponding spectra.

## Experimental DOS feature

To have a better understanding of the valence band from an experimental point of view, the low BE range was studied in more detail. In both spectra the baseline was corrected (see the Supp. Info. for background treatment) using the background function developed by Li et al.[45] As proposed by Fleming et al.,[40] for empirical reasons, the resulting spectra can be fitted with four components (Figure 3). Fleming's interpretation was deduced from UPS analysis on $TiO_2$/Si(100) grown by molecular beam deposition.[40]

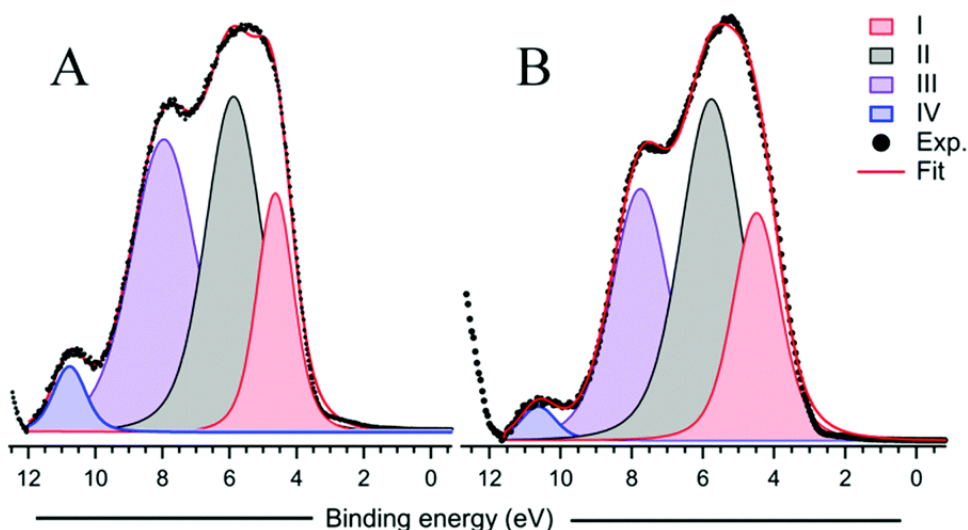

**Figure 3:** He I UP background subtracted spectra of anatase/ITO (drop method) (A) and rutile/ITO (EPD method) (B) and their decomposition into four asymmetric contributions.

The state observed at BE = 11 eV (IV) is assigned to surface OH groups.[60] Above this first contribution, decomposition to the remaining three states can be qualitatively explained on the basis of localized molecular orbital (MO) models,[63] the local electronic structure was associated with the MO-like state contributions from the Ti 3d and O 2p orbitals.

For titania, the molecular energy-level diagram of the valence band is composed of energy levels, due only to the non-bonding (π) O 2p orbitals ($t_{1g}$ and $t_{2u}$ levels). Then, the hybridization between the Ti orbitals with the O 2p orbitals forms, depending of their respective orientation, non-bonding orbitals ($3t_{1u}$ and $1t_{2g}$) and, deeper in energy, bonding orbitals (σ) ($2t_1u$, $2e_g$ and $2a_{1g}$).[64]

More recently, Van de Krol for rutile[65] and Asahi et al. for anatase[66] used MO theory to build an energy band diagram. The latter decomposed the VB into three main regions: the first one attributed to the O 2p π orbitals whereas π-hybridized and π-hybridized orbitals contribute to the second and the third regions. It was also correlated with DOS calculations. Indeed, UPS probes the DOS below the Fermi level and a better understanding can be obtained from the partial density of states (PDOS) description reported for anatase,[66,67] rutile,[55,68,69] both[23,24] or specific planes of titania.[70] Recently, DFT calculations on $TiO_2$ nanospheres corroborated the three state decomposition of UPS spectra.[71]

In the range between 0 and −6 eV, below the valence band maximum, DOS calculations predict three bands with a major contribution of O 2p orbitals. PDOS calculations clearly evidence that the contribution located at −1 eV below the $E_{VB}$ (state I) is due to O 2p π orbitals. Below this feature, the PDOS of Ti 3d increases with the decrease of energy (bonding states). State I is consequently due to a contribution from the hybridized non-bonding orbitals and state II is due to a contribution from the hybridized bonding orbitals.

### Band gap ($E_g$): UV-Vis diffuse reflectance spectroscopy

Overall, UPS accurately describes the occupied levels comprising the valence band. However, the positions of empty states are not probed by UPS. The band diagram can be completed to reach the level of the conduction band by using the data of UV-Vis DRS.

UV-Vis spectra are reported for the two phases in Figure S8. They show that, for a similar dilution in $BaSO_4$, the maximum of absorption is 2.5 higher for anatase than for rutile, due to the difference between the absorption coefficients of the two phases. It also demonstrates that using highly diluted samples is crucial to resolve fine transitions. Individual peaks could be resolved in the UV-Vis spectra of the 0.39 ppm diluted sample whereas for a more concentrated sample (16 ppm) this feature is hidden by peak broadening (Figure S8).

The decomposition of the t(hν) spectrum is illustrated in Figure 4. Four contributions are observed for the anatase sample and five for the rutile sample. For the anatase sample, an additional small sharp peak, probably related to excitons, at 3.25 eV is observed (the inset of Figure 4A, obtained with a dilution of 9.8 ppb).

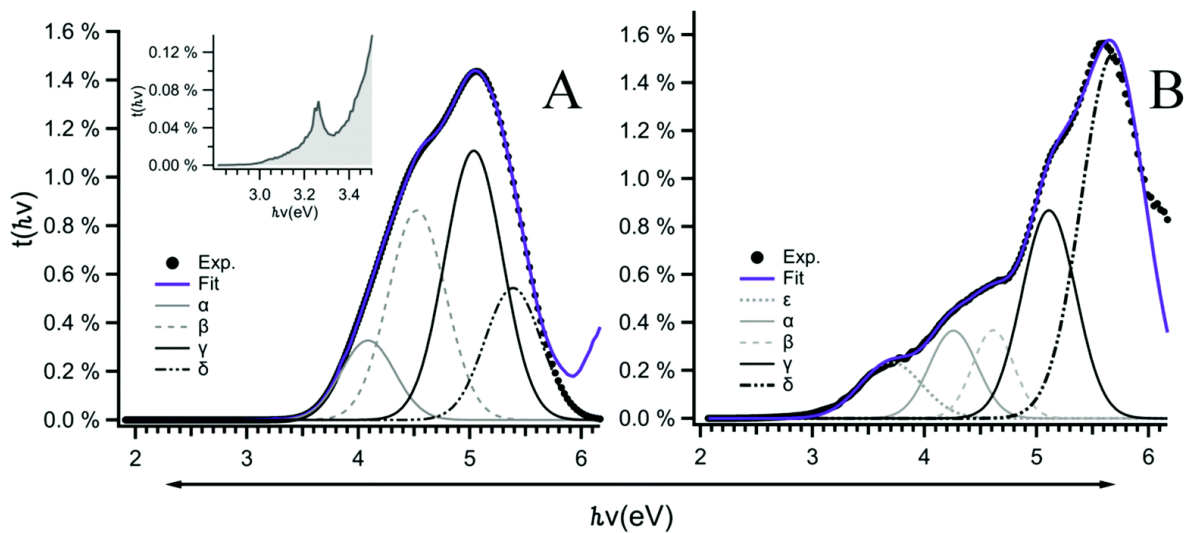

**Figure 4:** t(hν) versus energy spectra of $TiO_2$-anatase (A) and $TiO_2$-rutile (B) diluted with $BaSO_4$ (dilution of 0.39 ppm) (black dots). The violet curve is the fitting result and bell shape curves are the different contributions (α, β, γ, δ, and ε).

The four contributions around 4.2 eV (α), 4.5 eV (β), 5.0 eV (γ) and 5.5 eV (δ) are common to the two phases. A peak at lower energy is observed at 3.7 eV for rutile (ε). It may be due to a state near the $E_{VB}$ of rutile and nonexistent in anatase. For the two polymorphs the optical band gap was measured, $E_{g,anatase}$ = 3.15 eV and $E_{g,rutile}$ = 3.00 eV (see Figure S9). The error is estimated to be 0.05 eV.

## Absolute electronic structure of anatase and rutile

The relative positions of the VB determined from UPS and the $E_g$ value from UV-visible spectra give the overall picture of band positions corresponding to the staggered alignment in which anatase bands are lower than the rutile ones (type II-anatase), in agreement with the previous findings.[16,22,23] It must be noted that their $E_{VB}$ was determined separately, so a band alignment between the two contacting phases may shift those values.

To properly describe the electronic structures of both rutile and anatase, an absolute measurement of their energy levels was done. Dropping a suspension of $TiO_2$ in alkane on a silver substrate allowed measuring a UPS spectrum with both contributions from Ag and $TiO_2$ (Figure S7). Thus, on the same spectrum we observed the Fermi edge of the Ag foil and some contributions of its d-states as well as characteristic peaks belonging to $TiO_2$.

The VB maximum for the anatase phase is measured to be 2.9 eV below the Fermi level of silver. Additionally, the work function of this Ag foil was measured to be 4.2 eV. As a result, $E_{VB}$ for $TiO_2$ is −7.1 eV below the vacuum level. Using the offset between the $E_{VB}$ of the two polymorphs determined by UPS and using the two $E_g$ values determined by UV-Vis absorption, the $E_{CB}$ levels were deduced. The absolute electronic structure for both rutile and anatase is shown in Figure 5. To the best of our knowledge, for the first time, the band positions and shape of the VB of two powder photocatalysts were determined with such accuracy. A band offset of 0.4 eV for the valence band and 0.25 eV for the conduction band was determined between the two phases.

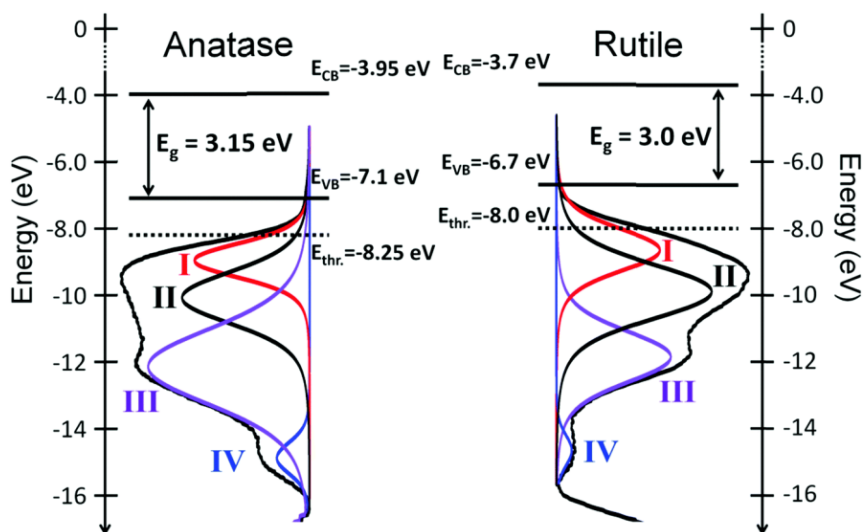

**Figure 5:** Electronic structure of the two $TiO_2$ anatase and rutile. $E_{thr.}$ corresponds to the deepest state that can be excited in a photocatalytic test ($E_{photons} \leq 4.3$ eV).

The reasons for this offset between the two polymorphs are both a shift of the different contributions and also the broadening of the O 2p state. Indeed, there is a 0.2 eV shift between the VB states of rutile and anatase (see Table S4). In addition, the full width at half maximum for contribution I is 1.3 eV for anatase and 1.6 eV for rutile. To achieve the MO-like description of UPS spectra, it was argued that close to the VB and CB edges, the local structure dominates over the effects of periodicity in determining the optical properties of $TiO_2$.[40] Such analysis allows a qualitative attribution of the corresponding components in the UP spectra of the valence band. However, it cannot be applied to explain the VB offset. Indeed, the local environment of Ti atoms in anatase and rutile is very similar and cannot account for the 0.4 eV differences of the band positions. Obviously, structure and symmetry related differences in the VB and CB band shape explain this energy offset. Due to such symmetry differences, rutile is a direct semiconductor with a transition at the G-point, whereas anatase is an indirect one.[72] The distribution of projected DOS in the anatase and rutile VB is similar. However, a simple qualitative interpretation of their band structure difference can be given. The valence bands primarily consist of O 2p states and a small Ti 3d contribution, indicating the presence of some p–d hybridization (which corresponds to component I in Figure 3). Hence, the difference between anatase and rutile is related to the O 2p-like states at the VB edge. Among two phases, rutile is a stable polymorph with a higher density and therefore a higher overall interaction between the Ti and O valent electronic manifolds. Due to this, the topmost VB band related to the O 2p states is broader in rutile which explains both the increase of VB edge energy and the decrease of $E_g$ as compared with the anatase counterpart. The bands in the CB are narrower since they are related to the antibonding states. The difference of CB properties is therefore lesser and the values for the two polymorphs are close.

We can now reconsider the UV-Vis electronic transitions observed in Figure 4 and assign them in terms of a simplified MO approach. Obviously α and β transitions (as well as e for rutile) are issued from non-bonding O 2p orbitals. Above, γ and δ transitions are energetic enough to photoexcite either non-bonding O 2p or hybridized orbitals.

In a typical standard photocatalytic test, Pyrex glassware filters the incoming photons below 290 nm ($E_{photons}$ = 4.3 eV).[73] This allows defining an energy threshold ($E_{thr.}$) corresponding to the deepest energy level above which the electrons can be photoexcited. This value is found to be −8.25 eV for anatase and −8.0 eV for rutile. Our work demonstrates that under UV illumination, electrons promoted from the VB to the CB come mainly from the non-bonding O 2p orbitals (contribution I) and contributed to photocatalysis. In order to optimize the photocatalytic efficiency, only those states must be targeted.

The type II-anatase alignment supported by our results suggests that photogenerated holes in anatase should have a higher oxidative potential than those in rutile and photogenerated electrons in rutile should have a higher reduction potential than those in anatase (Figure 1). The higher difference of oxidative potential counterbalances the difference of reduction potential and reinforces the conclusion that anatase is considered as the most active $TiO_2$ polymorph.[74]

## Conclusions

The powders of the two main polymorphs of $TiO_2$, anatase and rutile, were studied by UPS and UV-Vis spectroscopies. Up to now, UPS on the powders has been only qualitatively interpreted. It can provide a quantitative description of the energy levels comprising the valence band. Combination of UV-Vis and UPS allowed positioning of the energetic levels and discussion of the electronic structures of the two solids.

In the case of anatase and rutile powders the absolute $E_{VB}$ positions were measured at −7.1 eV and −6.7 eV, respectively. The resulting electronic structure obtained one at a time for anatase and rutile supports the hypothesis of a type II-anatase energy band alignment. In a mixture of both polymorphs the transfer of the electrons from rutile to anatase particles should be favoured. Considering the photocatalytic experiments, efficient photoelectrons for this process are excited from the top of the VB, attributed to a narrow contribution originating from the non-bonding O 2p orbitals.

A full description of the electronic structure of the powder samples would be obtained using complementary techniques such as inverse photoemission spectroscopy or spectroelectrochemical analysis to describe the CB states.[75,76]

## Conflicts of interest

There are no conflicts to declare.

## Acknowledgements

C. M. acknowledges financial support from the French Ministry of Education.